\documentclass[journal]{IEEEtran}
\ifCLASSINFOpdf
\else
\fi
\usepackage{amsmath}
\usepackage{graphicx}
\usepackage{subcaption}
\usepackage{multirow}

\begin{document}
%
\title{High-throughput Onboard Hyperspectral Image Compression with Ground-based CNN Reconstruction}
%
%
%

\author{Diego Valsesia,~\IEEEmembership{Member,~IEEE,}
        Enrico Magli,~\IEEEmembership{Fellow,~IEEE}
\thanks{The authors are with Politecnico di Torino -- Department of Electronics and Telecommunications, Italy. email: \{name.surname\}@polito.it. The research leading to this publication has received funding from the European Union's Horizon 2020 research and innovation programme under grant agreement No 776311.}}

\maketitle

\begin{abstract}
Compression of hyperspectral images onboard of spacecrafts is a tradeoff between the limited computational resources and the ever-growing spatial and spectral resolution of the optical instruments. As such, it requires low-complexity algorithms with good rate-distortion performance and high throughput. In recent years, the Consultative Committee for Space Data Systems (CCSDS) has focused on lossless and near-lossless compression approaches based on predictive coding, resulting in the recently published CCSDS 123.0-B-2 recommended standard. While the in-loop reconstruction of quantized prediction residuals provides excellent rate-distortion performance for the near-lossless operating mode, it significantly constrains the achievable throughput due to data dependencies. In this paper, we study the performance of a faster method based on prequantization of the image followed by a lossless predictive compressor. While this is well known to be suboptimal, one can exploit powerful signal models to reconstruct the image at the ground segment, recovering part of the suboptimality. In particular, we show that convolutional neural networks can be used for this task and that they can recover the whole SNR drop incurred at a bitrate of 2 bits per pixel.    
\end{abstract}


%
\IEEEpeerreviewmaketitle

\section{Introduction}
Hyperspectral imaging from spaceborne spectrometers enables a wide range of applications, including material identification, terrain analysis and military surveillance. The ever-increasing spectral and spatial resolution of such instruments allows to create higher and higher quality products for the final user but it poses challenges in handling such wealth of data. In particular, onboard compression is critical to overcome the limited downlink bandwidth. This area of research poses specific challenges due to the strict complexity limitations on the payload hardware. Several solutions based on different techniques have been proposed, such as low-complexity spatial \cite{ccsds122} and spectral transforms \cite{pot}, distributed source coding \cite{abrardoDSC}, compressed sensing \cite{universal,valsesia2016vector,barducci2014compressive}, and predictive coding \cite{acap,mcalic,calibrationartifacts,ccsds123}. Predictive coding has emerged as one of the most popular solutions, as it enables low-complexity, high-throughput solutions, excellent rate-distortion performance and flexibility in the definition of image quality policies \cite{hydra,valsesia_icip,valsesia_letter,conoscenti}. The CCSDS has been working on extending the CCSDS 123.0-B-1 recommendation \cite{ccsds123} for predictive lossless compression, resulting in the recent publication of the 123.0-B-2 recommendation \cite{ccsds123nl}. The new standard extends the previous one in the lossless mode, and includes lossy compression modes based on the introduction of a quantizer and a local decoder inside the prediction loop. It is well known \cite{jayant1984digital} that an in-loop quantizer provides better rate-distortion performance than quantization followed by lossless predictive coding. However, one must consider that the need for a local decoder to reconstruct pixel values in the prediction neighborhood creates data dependencies which prevent parallelization and, consequently, high-throughput operations. 

Meanwhile, recent years have seen the rise of neural networks as data-driven methods to solve problems previously tackled with hand-crafted models. In particular, imaging problems have been revolutionized by convolutional neural networks (CNNs). CNNs are able to capture very complex models about natural images because the convolution operation exploits powerful image priors such as shift invariance, and compositionality, where a complex global model is constructed from nonlinear hierarchies of local features. Ultimately, CNNs have proved to be able to achieve state-of-the-art performance on a wide variety of tasks including classification \cite{hu2018squeeze}, segmentation \cite{kaiser2017learning}, object detection \cite{yolo} and regularization of inverse problems such as denoising \cite{zhang2017beyond} and superresolution \cite{dong2016image, lei2017superresolution, BordoneWHISPERS19}. 

In this paper, we propose to combine a low-complexity onboard compressor of hyperspectral images with a CNN-based reconstruction algorithm working at the ground segment. The main objective is to study its the rate-distortion performance with respect to the latest CCSDS standard. It is known that midpoint reconstruction from quantized data is not always optimal for image reconstruction, and e.g., using uniform-threshold quantization and a Laplacian assumption on the residuals is better. CNNs do not require an a priori model of the residuals, but are able to learn this model from training data. We show that the CNN learns to exploit the spatial and spectral correlation patterns of natural images to regularize the inverse reconstruction problem, and can be very effective at improving the quality of the image. Armed with such a powerful tool that runs at the ground segment where computational resources are abundant, one may wonder how much complexity is really needed onboard where resources are scarce. Preliminary FPGA implementations of the CCSDS 123.0-B-2 standard (using the Golomb entropy encoder) show that the lossless algorithm can achieve throughputs in excess of 100 Msamples/s \cite{fpga123_1,shyloc}, while its lossy counterpart is limited to 20 Msamples/s \cite{denino2014lossy} due to the aforementioned data dependencies. The new standard addresses this issue with a coding mode dedicated to high-throughput scenarios by removing some data dependencies, at a cost in terms of rate-distortion performance. In this paper, we propose to replace the lossy standard compressor with a different scheme based on prequantization of the raw pixels followed by the lossless CCSDS 123.0-B-2 encoder and a CNN reconstructor at the ground segment. The throughput of this compressor is essentially limited by the lossless predictor which is fast due to the lack of data dependencies. We show that the suboptimality due to moving the quantizer outside the prediction loop can be fully recovered by the CNN reconstruction and the same rate-distortion performance as lossy CCSDS 123.0-B-2 (without the CNN) is achieved, while potentially achieving the same throughput of the lossless version of the recommendation.

A preliminary version of this work appeared in \cite{obdp}. With respect to the conference version, the method and its analysis are more thoroughly explained, we expand the treatment by also considering a relative error objective, present new experiments on a larger test set, and discuss transfer learning to different sensors. 
The paper is organized as follows. Sec. \ref{sec:bkg} provides some background material on the CCSDS 123.0-B-2 recommendation for lossy compression. Sec. \ref{sec:dequantization} details the CNN used for image reconstruction. Sec. \ref{sec:compression} outlines the two approaches to onboard compression analyzed in the paper, i.e., lossy CCSDS 123.0-B-2 and prequantization followed by lossless CCSDS 123.0-B-2, for two quality objectives, namely bounded absolute or relative error. Sec. \ref{sec:experiments} discusses the experimental results. Finally, Sec. \ref{sec:conclusions} draws some conclusions.

\section{Background on CCSDS 123.0-B-2} \label{sec:bkg}
The CCSDS issued the Blue book for the 123.0-B-1 recommendation in May 2012 \cite{ccsds123} and an Issue 2 in February 2019 \cite{ccsds123nl}. The original recommendation focused on defining a method for lossless compression of hyperspectral images based on predictive coding. In particular, it is based on the fast lossless \cite{FL} predictor, which uses an adaptive filter to estimate a pixel value from information in a causal neighborhood. The prediction residual is then entropy coded by means of Golomb power of 2 (GPO2) codes \cite{golomb1966run}. This recommendation has been recently subject to a revision in order to extend it to lossy compression, resulting in the CCSDS 123.0-B-2 standard \cite{ccsds123nl}. This extension is essentially based on the near-lossless coding principle, whereby a prediction residual, i.e., the difference between the predicted and the original pixel values, is quantized and locally decoded in order to update the weights of the prediction filter with the sign algorithm \cite{sign}. The extended recommendation also introduces a new prediction mode, namely \emph{narrow local sums}, which essentially avoids using the pixel immediately on the left and in the same band of the pixel being coded. This mode is motivated by reasons of implementation efficiency: due to the local decoder in the prediction loop, the current pixel cannot be predicted unless every pixel in the causal neighborhood under consideration has already been coded and decoded. The pixel on the left is especially important because it is coded immediately before the current one in the popular BSQ and BIL orderings and it is the main bottleneck in hardware implementations. 

More in detail, the algorithm computes a local sum $\sigma_{x,y,z}$ which is defined as 
\begin{align*}
\sigma_{x,y,z} = \begin{cases}
s^{R}_{x-1,y,z} + s^{R}_{x-1,y-1,z} + s^{R}_{x,y-1,z} + s^{R}_{x+1,y-1,z}, \quad \hfill \\ \hfill y>0,0<x<N_x-1 \\
4 s^{R}_{x-1,y,z}, \hfill y=0,x>0 \\
2(s^{R}_{x,y-1,z}+s^{R}_{x+1,y-1,z}), \hfill y>0,x=0 \\
s^{R}_{x-1,y,z} + s^{R}_{x-1,y-1,z} + 2 s^{R}_{x,y-1,z}, \\ \hfill y>0,x=N_x-1
\end{cases}
\end{align*}
for the wide, neighbor-oriented mode and as
\begin{align*}
\sigma_{x,y,z} = \begin{cases}
s^{R}_{x-1,y-1,z} + 2s^{R}_{x,y-1,z} + s^{R}_{x+1,y-1,z},\\ \hfill y>0,0<x<N_x-1 \\
4 s^{R}_{x-1,y,z-1}, \hfill y=0,x>0 \\
2(s^{R}_{x,y-1,z}+s^{R}_{x+1,y-1,z}), \hfill y>0,x=0 \\
2(s^{R}_{x-1,y-1,z} + s^{R}_{x,y-1,z}), \qquad \hfill y>0,x=N_x-1 \\
4 s_{mid}, \hfill y=0,x>0,z=0
\end{cases}
\end{align*}
for the narrow, neighbor-oriented mode, being $s^{R}_{x,y,z}$ the reconstructed pixel at position $(x,y,z)$. Column-oriented modes also exist but will not be considered in this paper, as they are mostly intended for images with striping artifacts.
The reduced prediction mode only uses the central local difference $d_{x,y,z}=4s^{R}_{x,y,z}-\sigma_{x,y,z}$ while the full prediction mode also uses directional local differences $d^N_{x,y,z}$,$d^W_{x,y,z}$,$d^{NW}_{x,y,z}$ (we refer the reader to \cite{ccsds123nl} for more details on the definitions). The predicted central difference $\hat{d}_{x,y,z}$ is obtained by multiplying the adaptive filter weights with the vector of differences, i.e.,
\begin{align*}
\hat{d}_{x,y,z}=\mathbf{W}_{x,y,z}\left[\begin{array}{c}
d_{x,y,z}^{N}\\
d_{x,y,z}^{W}\\
d_{x,y,z}^{NW}\\
d_{x,y,z-1}\\
d_{x,y,z-2}\\
\vdots\\
d_{x,y,z-P}
\end{array}\right]
\end{align*}
for full mode and 
\begin{align*}
\hat{d}_{x,y,z}=\mathbf{W}_{x,y,z}\left[\begin{array}{c}
d_{x,y,z-1}\\
d_{x,y,z-2}\\
\vdots\\
d_{x,y,z-P}
\end{array}\right]
\end{align*}
for reduced mode.
The predicted central difference is then transformed to obtain the predicted pixel value $\hat{s}_{x,y,z}$.

Finally, the recommendation also provides new tools such as sample representatives and new hybrid entropy coder able to reach rates lower than 1 bit per pixel (bpp), overcoming the limit of the original GPO2 encoder. Two main objectives can also be specified to drive the in-loop quantizer: bounded absolute error or bounded relative error.

\begin{figure*}[h]
    \centering
    \includegraphics[width=0.9\textwidth]{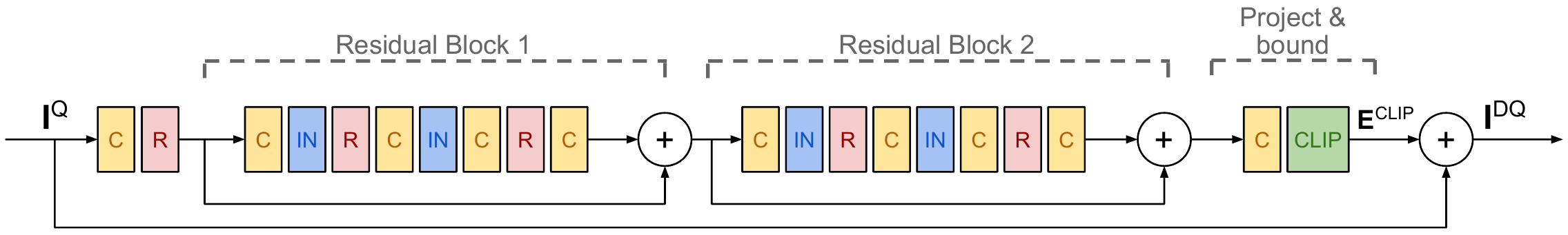}
    \caption{Reconstruction CNN. C: 2D convolution, R: leaky ReLU, IN: 2D instance normalization, CLIP: residual clipping. Input and output sizes are $N_l \times N_c \times 8$.}
    \vspace{-0.5cm}
    \label{fig:dequant_net}
\end{figure*}

\section{Reconstruction using convolutional neural networks} \label{sec:dequantization}

This section presents the proposed approach to recover part of the image information lost during the lossy compression process. Any kind of lossy compression introduces artifacts which change the distribution of pixel values with respect to the one exhibited by natural uncompressed images. Recovering the original image from its distorted version is an ill-posed inverse problem, as there are infinitely many solutions. However, it is possible to compute a better estimate of the original image by properly modelling what constitutes a natural image.

Traditional techniques relied on hand-crafted image priors to model image data. For instance, a popular technique is total variation minimization, which amounts to requiring that the energy of the gradients in a natural image should be small. Image recovery from a compressed image $\mathbf{I}^Q$ is cast as the solution to the following minimization problem:
\begin{align}
\label{eq:tv_min}
    \mathbf{I}^{DQ} = \arg&\min_\mathbf{I} \Big[ \Vert \mathbf{I} - \mathbf{I}^Q \Vert^2_2 +\lambda \sum_{x,y,z}( \vert I_{x+1,y,z} - I_{x,y,z} \vert + \nonumber \\&+ \vert I_{x,y+1,z} - I_{x,y,z} \vert + \vert I_{x,y,z+1} - I_{x,y,z} \vert ) \Big].
\end{align}

Recently, convolutional neural networks (CNNs) have shown remarkable results in a variety of inverse problems, including denoising and superresolution. Their success lies in their ability to create more sophisticated models of complex image data as well as being able to handle perturbations with non-trivial statistics (e.g., non-Gaussian).

\vspace{-0.1cm}
\subsection{Proposed CNN}
\label{sec:cnn}

The proposed CNN reconstructs a better estimate of the original image from decoded hyperspectral images after lossy compression. Its training objective is to minimize the mean squared error (MSE) between the reconstructed image and the original. It is important to notice that the reconstruction depends on the specific algorithm used for compression and also the chosen quality level. This is similar to the denoising problem where several algorithms are based on knowing the noise variance \cite{zhang2017beyond,lehtinen2018noise2noise}. In our case, we train a CNN to invert a specific compression algorithm (e.g., near-lossless CCSDS 123.0-B-2) at a specific quality point which is known from the compression system design (e.g., a fixed quantizer step size for bounded absolute error near-lossless compression). We also argue that the trained model is optimal for new images acquired by the same sensor, as the network learns to exploit the peculiar spatial and spectral correlation patterns produced by that sensor. Nevertheless, the CNN has some generalization capability to unseen sensors as some feature extraction steps are common for all sensors, thus only requiring fine-tuning with a smaller amount of data. Concerning the MSE training loss, some works have addressed image restoration using adversarial losses \cite{divakar2017image,chen2018image}, i.e., a game between two networks, one restoring the image, the other discriminating whether its input is an original or restored image. We will not consider this kind of loss because it tends to hallucinate image details which might be visually pleasing \cite{ledig2017photo}, but not really part of the original image and, in fact, such objective typically yields higher MSE values.

Fig. \ref{fig:dequant_net} shows an overview of the network. The input to the network is a slice of a hyperspectral image of size $N_l \times N_c \times 8$, where $N_l$ and $N_r$ are the number of lines and columns, respectively. While the spatial dimensions can be arbitrary, the number of bands is fixed to 8 in our proposed design. The main reason for this choice is the use of two-dimensional convolutional layers instead of three-dimensional ones. The first convolutional layer of the network has 64 filters of size $3 \times 3 \times 8$, thus merging the information from the 8 bands without sliding the kernel in the spectral dimension. A three-dimensional convolutional layer would have had a sliding kernel over all the three dimension and would have allowed an arbitrary number of spectral channels in the input. However, we found two main issues with this approach: \textit{i)} the large size of hyperspectral images calls for careful memory usage and 3D convolutions require a very large amount of memory; \textit{ii)} after reducing the use of memory to an acceptable value we found training to be highly unstable and providing results worse than those of the architecture with 2D convolutions. This is also an important design point in order to deal efficiently with images of large size. Notice that having a fixed number of input bands does not mean that only images with 8 bands can be processed. In fact, it is sufficient to slide a window over the spectral dimension of an image with more bands to process each slice and then merge the results. If partially overlapping slices are processed then the results are averaged by weighing each band by the number of times it has gone through the network.

The global input-output residual connection in the architecture means that the network learns to estimate the perturbation of the input image. This is an established solution in the literature on denoising \cite{zhang2017beyond}, as it allows solving a simpler task by removing low-frequency content predicted by the input image. The inner layers of network show two main residual blocks composed of alternating convolutions, instance normalization layers and leaky ReLU nonlinearities \cite{xu2015empirical}. The use of residual blocks was introduced by the ResNet architecture \cite{he2016resnet} for image classification and has multiple benefits such as reducing the vanishing gradient problem thanks to one of the addends skipping several layers and improved learning capability due to the need to only learn the residual of an identity mapping instead of the full mapping. Instance normalization \cite{vedaldi2016instance} normalizes activations to be approximately zero mean and unit standard deviation but, contrary to batch normalization \cite{ioffe2015batchnorm}, has different normalization factors for each image in the batch. Intuitively, this acts as a ``contrast normalization'' across the batch and helps dealing with perturbations that have more complex statistics than Gaussian noise, such as the case for reconstruction of compressed images.

Finally, the last layer allows to enforce consistent reconstruction, i.e., it ensures that the reconstructed pixel values fall in the same quantization bins as the original pixels by clipping the values of the correction estimated by the neural network. This is a design point that is specific to the reconstruction problem presented in this paper and also depends on the choice of the quantizer in the compression algorithm. In order to understand this, let us study a simple example. Suppose that the compression algorithm consists of simple uniform scalar quantization of the integer pixel values, i.e. $\mathbf{I}^Q = Q \lfloor \frac{\mathbf{I}}{Q} + \frac{1}{2} \rfloor$, with $Q=2\Delta+1$ for some integer $\Delta$. Then, we know that the error is bounded as $\vert\mathbf{I}^Q-\mathbf{I}\vert \leq \Delta$. If we call $\mathbf{E}^{\text{CLIP}}$ the correction term estimated by the network, then it must obey $\vert \mathbf{E}^{\text{CLIP}} \vert \leq \Delta$ since we know that the quantized pixel is never further than $\Delta$ from the original. Also notice that the bound on maximum error on the reconstructed image is, inevitably, twice the original bound.  

We want to emphasize that proposing an entirely novel CNN architecture is outside the scope of this paper. Instead, we are interested in assessing how a baseline design inspired by recent results in the literature can already show that the proposed approach is competitive. Further optimization is certainly possible, e.g., by exploiting non-local features \cite{N3Net, ValsesiaICIP19}. However, this further strengthens the main point of this paper, which is about showing that coupling a simpler on-board compressor with a CNN at the ground segment allows higher throughput and has competitive rate-distortion performance with respect to the lossy CCSDS 123.0-B-2 standard.

\section{Onboard compression approaches}
\label{sec:compression}

\begin{figure}[t]
    \centering
    \begin{subfigure}{0.48\textwidth}
    \includegraphics[width=\textwidth]{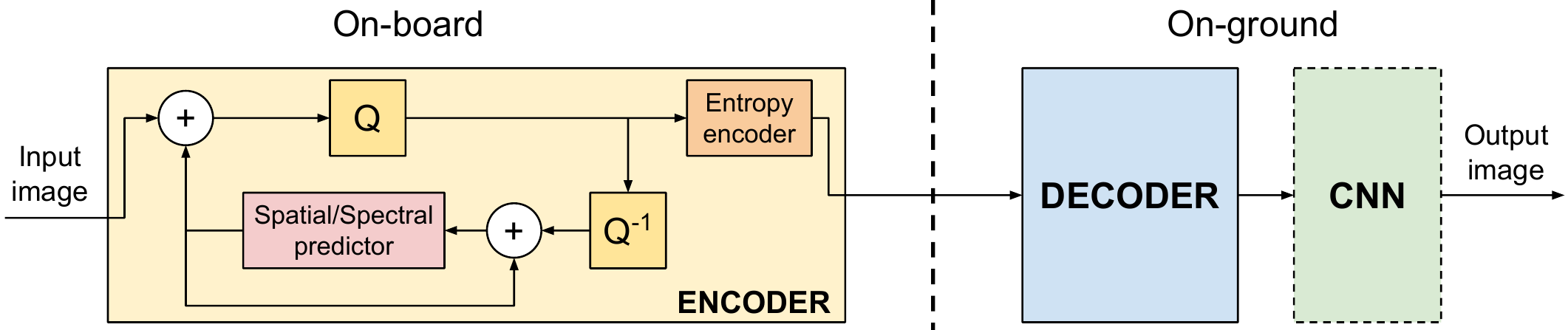}
    \caption{CCSDS 123.0-B-2 lossy compressor.}
    \end{subfigure}
    \begin{subfigure}{0.48\textwidth}
    \includegraphics[width=\textwidth]{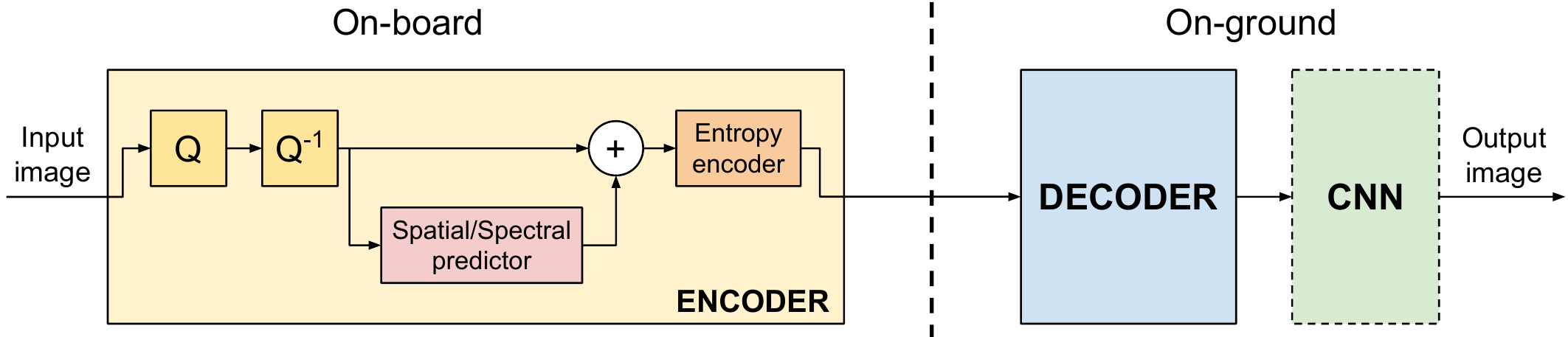}
    \caption{Prequantization lossy compressor.}
    \end{subfigure}
    \caption{Two predictive compression approaches. CCSDS 123.0-B-2 uses a quantizer inside the prediction loop. Prequantization quantizes raw pixel data and then applies a lossless predictor.}
    \vspace{-0.5cm}
    \label{fig:compressors}
\end{figure}

This section discusses two approaches to lossy onboard compression of hyperspectral images, namely the new CCSDS 123.0-B-2 recommendation and a simpler algorithm based on scalar quantization of the pixel values followed by a lossless predictive coding scheme, which we choose to be the lossless mode of CCSDS 123.0-B-2. We will refer to this method as ``prequantization''. Fig. \ref{fig:compressors} visually depicts the two methods. We study the performance of both algorithms for two quality objectives: bounded absolute error and bounded relative error. We also study the performance impact of an on-ground reconstruction stage using the CNN presented in the previous section.

\subsection{Complexity and data dependencies}
The main reason to compare the two methods is to assess the most efficient way to employ the revised recommendation for lossy hyperspectral image compression. Scenarios requiring high-throughput implementations are particularly interesting, whereby the in-loop quantizer significantly limits the CCSDS algorithm. Recalling the notation of Sec. \ref{sec:bkg}, let us consider the wide, neighbor-oriented coding mode of lossy CCSDS 123.0-B-2 under band interleaved by line (BIL) coding order. The computation of the current local sum $\sigma_{x,y,z}$ requires knowing the value of $s^{R}_{x-1,y,z}$, i.e., the reconstructed pixel value on the left of the current pixel in the same band. In the BIL order, the $(x-1,y,z)$ pixel is coded immediately before the $(x,y,z)$ pixel, which implies that all computations for $(x-1,y,z)$ must be terminated before starting coding $(x,y,z)$. This prevents building efficient parallel pipelines where the computation of the local sum can be started for several pixels ahead of the one being coded. The lossless version of CCSDS 123.0-B-2 does not suffer from such dependency as it only requires the original pixel values, not the reconstructed ones. In fact, space-grade FPGA implementations \cite{fpga123_1,fpga123_2} of the lossless algorithm achieved a throughput in excess of 100 Msamples/s while a comparable FPGA implementation of the lossy standard \cite{denino2014lossy} was only limited to 20 Msamples/s due to this dependency issue.

The prequantization approach removes the quantizer from the prediction loop and therefore does not suffer from the same bottleneck. The prediction loop is lossless and can therefore achieve very high throughput, while the prequantization operation of the input data has negligible complexity compared to the predictor. Therefore, the prequantization method essentially shifts part of the complexity from the on-board encoder to the CNN needed after the decoder at the ground segment in order to recover the sub-optimal rate-distortion performance compared to the in-loop quantizer. The ground segment has fewer complexity issues and the main limitation is the memory usage of the GPU while reconstructing the image. This is limited by the design in Sec. \ref{sec:cnn} which uses 2D convolutions instead of more expensive 3D convolutions. The memory required by each 2D convolutional layer is $N_x N_y F$ floating point values instead of $N_x N_y N_z F$ floating point values required by 3D convolutions, being $F$ the number of layer filters (64 in our design) and $N_x \times N_y$ the spatial dimensions of the input image.

\vspace{-0.2cm}
\subsection{Bounded absolute error}
A guarantee bounding the absolute error is achieved by both the CCSDS and the prequantization methods by using a uniform scalar quantizer. In the former case, the quantizer operates on the prediction residuals, while in the latter case it is directly applied to the pixel values.

\subsection{Bounded Relative error}

\begin{figure}[t]
    \centering
    \includegraphics[width=0.68\columnwidth]{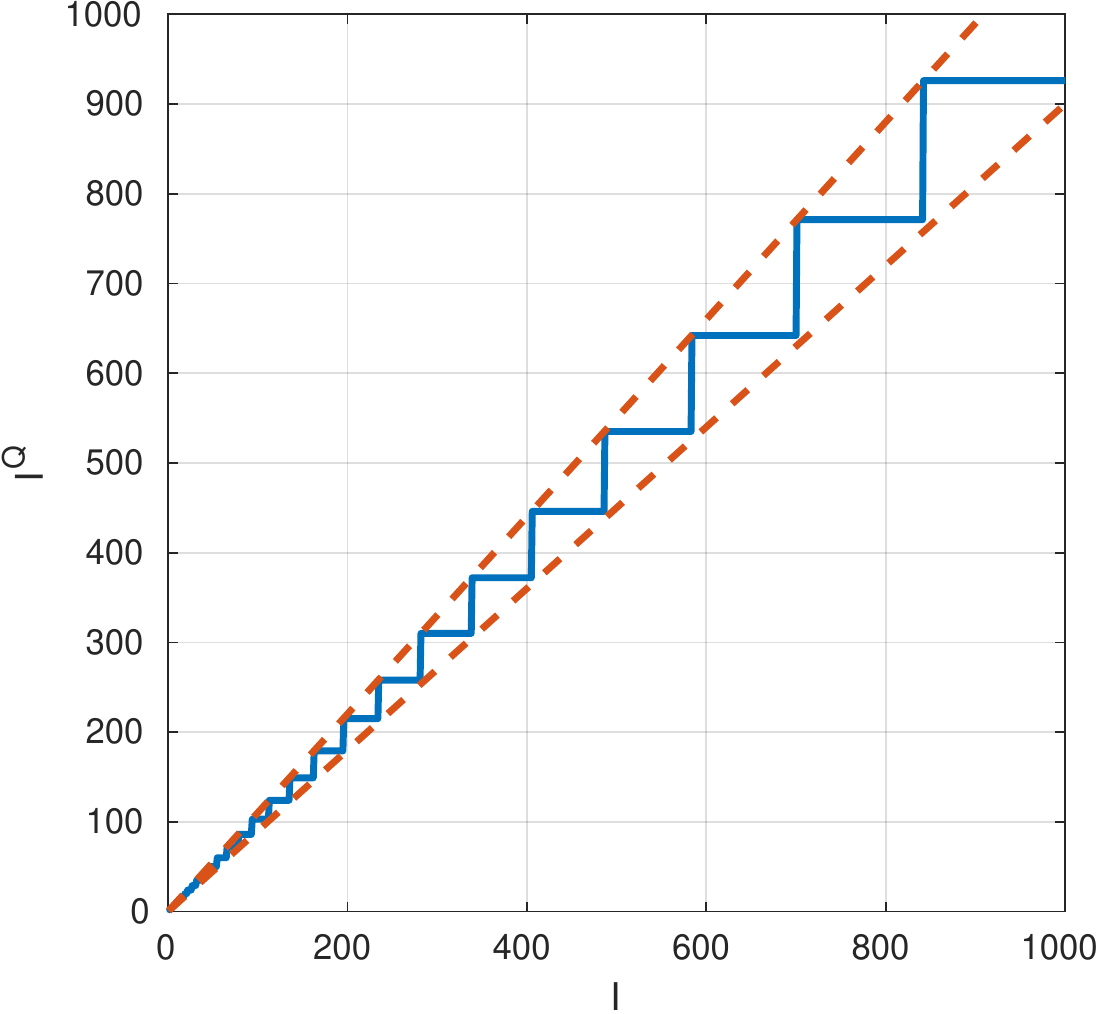}
    \vspace{-0.05cm}
    \caption{Relative error quantizer for prequantization method. Dashed lines show the $\pm 10\%$ error bound.}
    \vspace{-0.6cm}
    \label{fig:relerr_quant}
\end{figure}

A method to compress hyperspectral images using the CCSDS 123.0-B-2 standard with a target on relative error, rather than absolute error has been first proposed by Conoscenti et al. \cite{conoscenti} and it is included in the revised recommendation. The main idea is to use an in-loop uniform scalar quantizer whose quantization step size changes at every pixel as it depends on the predicted pixel value to approximate the desired relative error. In particular, the following formula is used:
\begin{align*}
    Q = 2 \lfloor R \vert \hat{s}_{x,y,z} \vert \rfloor + 1,
\end{align*}
being $R$ the target relative error and $\hat{s}_{x,y,z}$ the predicted pixel value.
Notice that the predicted pixel value is used rather than the original pixel value in order to maintain causal decodability. This does not provide a hard bound on the relative error, but the use of a safety margin in the formula to compute the desired quantization step size showed good performance, with rare instances of error beyond the chosen limit.

It is obvious that the prequantization method can achieve a bounded relative error guarantee by designing a non-uniform scalar quantizer, where large pixel values are more coarsely quantized according to the desired relative error. Fig. \ref{fig:relerr_quant} shows a sample design \cite{relerrccsds} of such quantizer, obtained by successive greedy extension of each quantization interval to match the relative error constraint.

\begin{figure*}[t]
    \centering
    \begin{subfigure}{0.48\textwidth}
    \includegraphics[width=\textwidth]{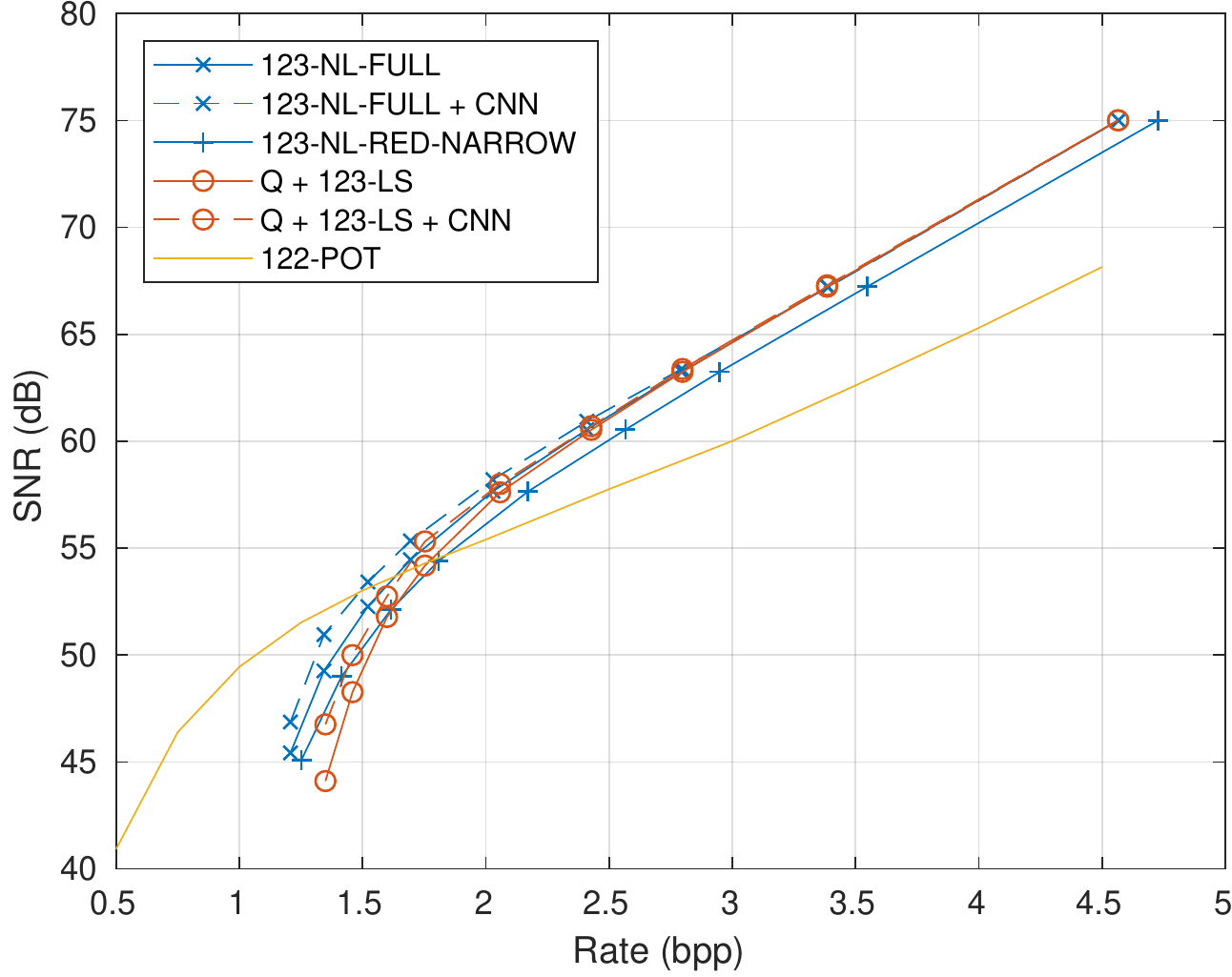}
    \caption{sc0}
    \end{subfigure}
    \begin{subfigure}{0.48\textwidth}
    \includegraphics[width=\textwidth]{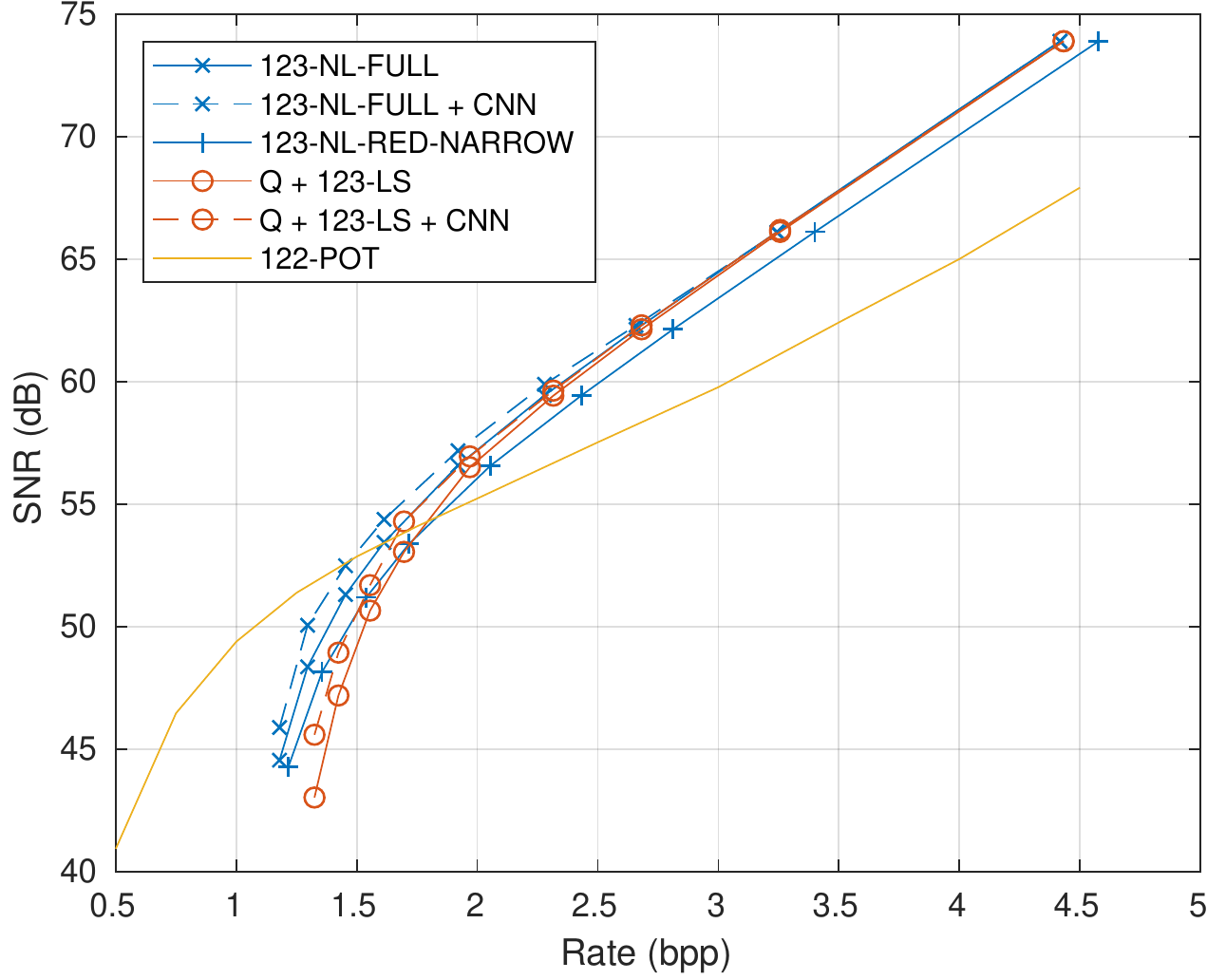}
    \caption{sc3}
    \end{subfigure}
    \begin{subfigure}{0.48\textwidth}
    \includegraphics[width=\textwidth]{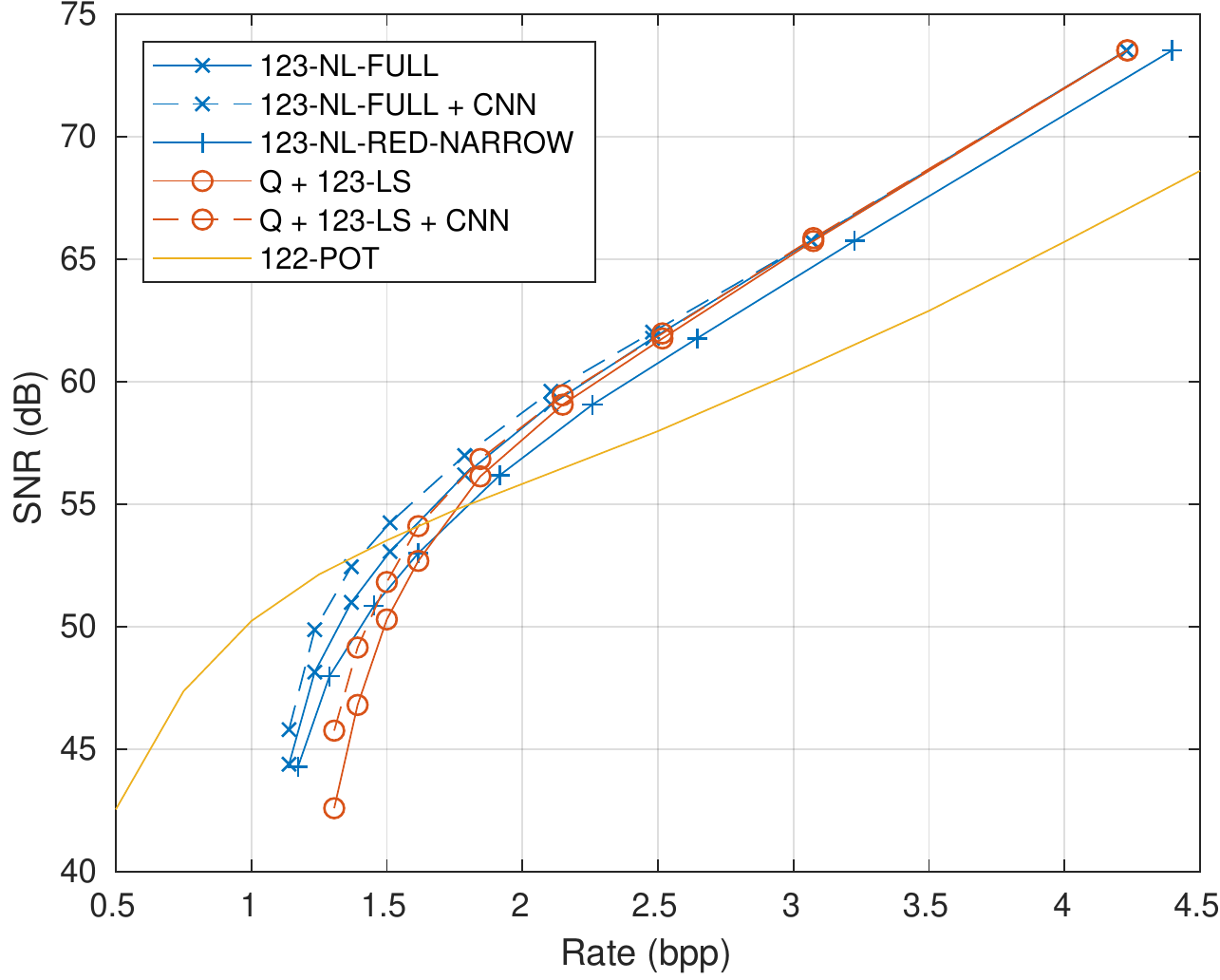}
    \caption{sc11}
    \end{subfigure}
    \begin{subfigure}{0.48\textwidth}
    \includegraphics[width=\textwidth]{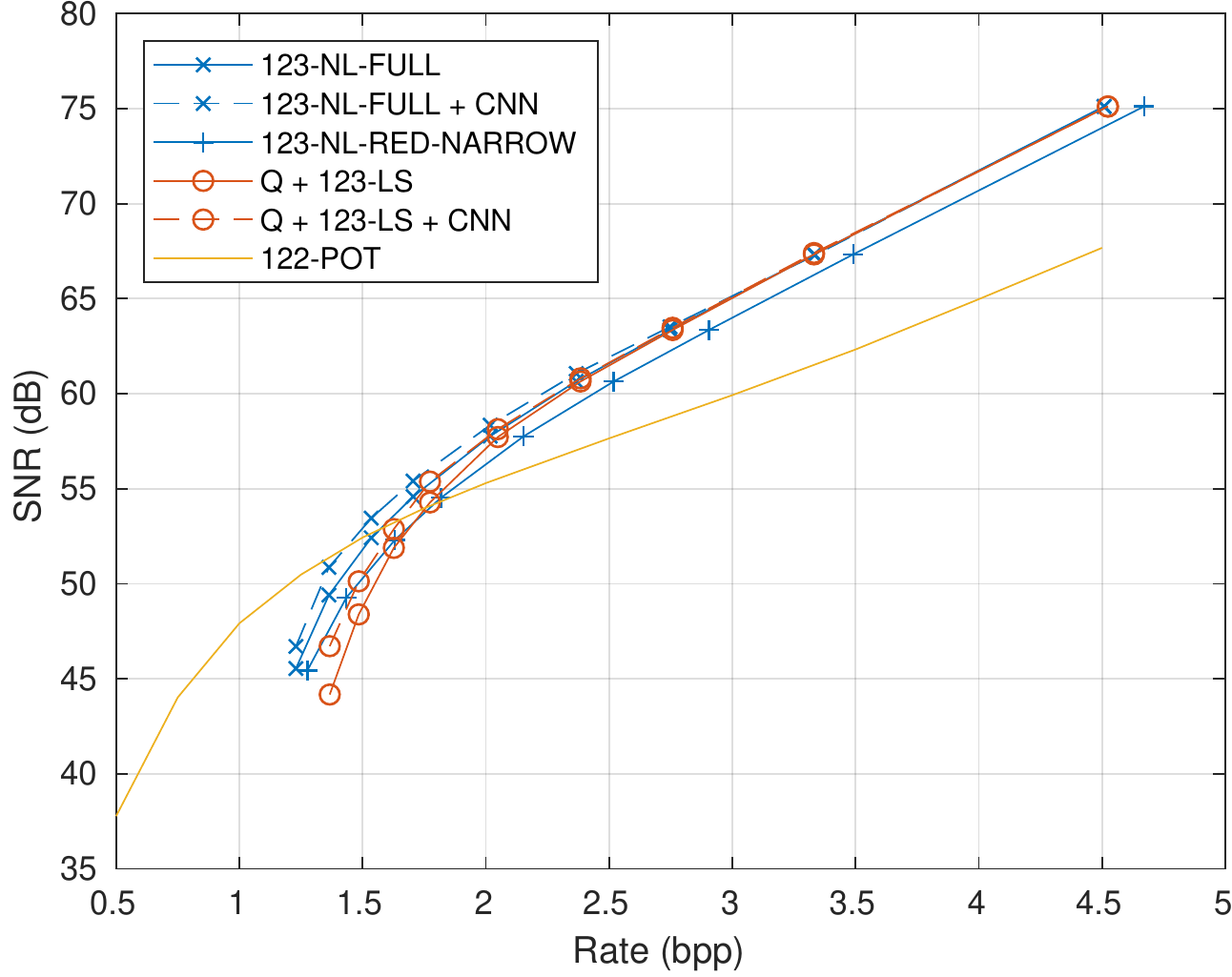}
    \caption{sc18}
    \end{subfigure}
    \caption{Rate-SNR performance of various compression methods with and without onground CNN. 123-NL: lossy CCSDS 123.0-B-2 (full, wide, neighbor-oriented mode); Q+123-LS: prequantization followed by lossless CCSDS 123.0-B-2 (full, wide, neighbor-oriented mode); 123-NL-RED-NARROW: lossy CCSDS 123.0-B-2 (reduced, narrow, neighbor-oriented mode); 122-POT: CCSDS 122 and POT; CNN: CNN reconstruction.}
    \vspace{-0.2cm}
    \label{fig:snr_bae}
\end{figure*}

\section{Experiments} \label{sec:experiments}
This section presents an experimental assessment of the performance of the proposed CNN reconstruction when combined with the two compression approaches presented in Sec. \ref{sec:compression}. For both approaches we set the CCSDS predictor in its full prediction mode with wide neighbor-oriented local sums. Their rate-distortion performance is measured against a number of baseline methods. A first baseline is a transform-coding approach to onboard hyperspectral image compressor where the CCSDS 122 recommendation \cite{ccsds122} for spatial compression using wavelets is combined with the Pairwise Orthogonal Transform (POT) to remove spectral correlation \cite{pot}. Another comparison is drawn with the CCSDS lossy compressor set in reduced prediction mode with narrow neighbor-oriented local sums. This is the recommended mode of the CCSDS standard to achieve high throughput at the expense of some compression performance.

\subsection{CNN training and testing details}

\begin{table*}[]
\normalsize
\centering
\caption{SNR (dB) for test set}
\label{table:snr}
\begin{tabular}{ccccccc}
                 & \textbf{123-NL}  & \textbf{123-NL + CNN} & \textbf{Q + 123-LS} & \textbf{Q + 123-LS + CNN} & \hspace{-9pt} \textbf{123-NL-RED-NARROW} \hspace{-9pt} & \textbf{122-POT} \\ \hline
\textbf{1.5 bpp} & $52.23 \pm 0.49$ & $\mathbf{53.38} \pm 0.50$      & $49.58 \pm 0.69$    & $51.10 \pm 0.73$          & $50.85 \pm 0.64$           & $53.13 \pm 0.54$ \\ \hline
\textbf{2.0 bpp} & $57.60 \pm 0.34$ & $\mathbf{58.19} \pm 0.36$      & $57.15 \pm 0.33$    & $57.65 \pm 0.36$          & $56.36 \pm 0.33$           & $55.53 \pm 0.31$ \\ \hline
\textbf{3.0 bpp} & $64.88 \pm 0.34$ & $\mathbf{64.93} \pm 0.33$      & $64.80 \pm 0.34$    & $64.92 \pm 0.35$          & $63.81 \pm 0.32$           & $60.09 \pm 0.28$ \\ \hline
\textbf{4.0 bpp} & $71.57 \pm 0.36$ & $\mathbf{71.55} \pm 0.36$      & $71.53 \pm 0.37$    & $71.56 \pm 0.36$          & $70.49 \pm 0.34$           & $65.33 \pm 0.36$  \\ \hline
\end{tabular}
\vspace{-0.1cm}
\end{table*}

\begin{figure}[t]
    \centering
    \begin{subfigure}{0.49\columnwidth}
    \includegraphics[width=\columnwidth]{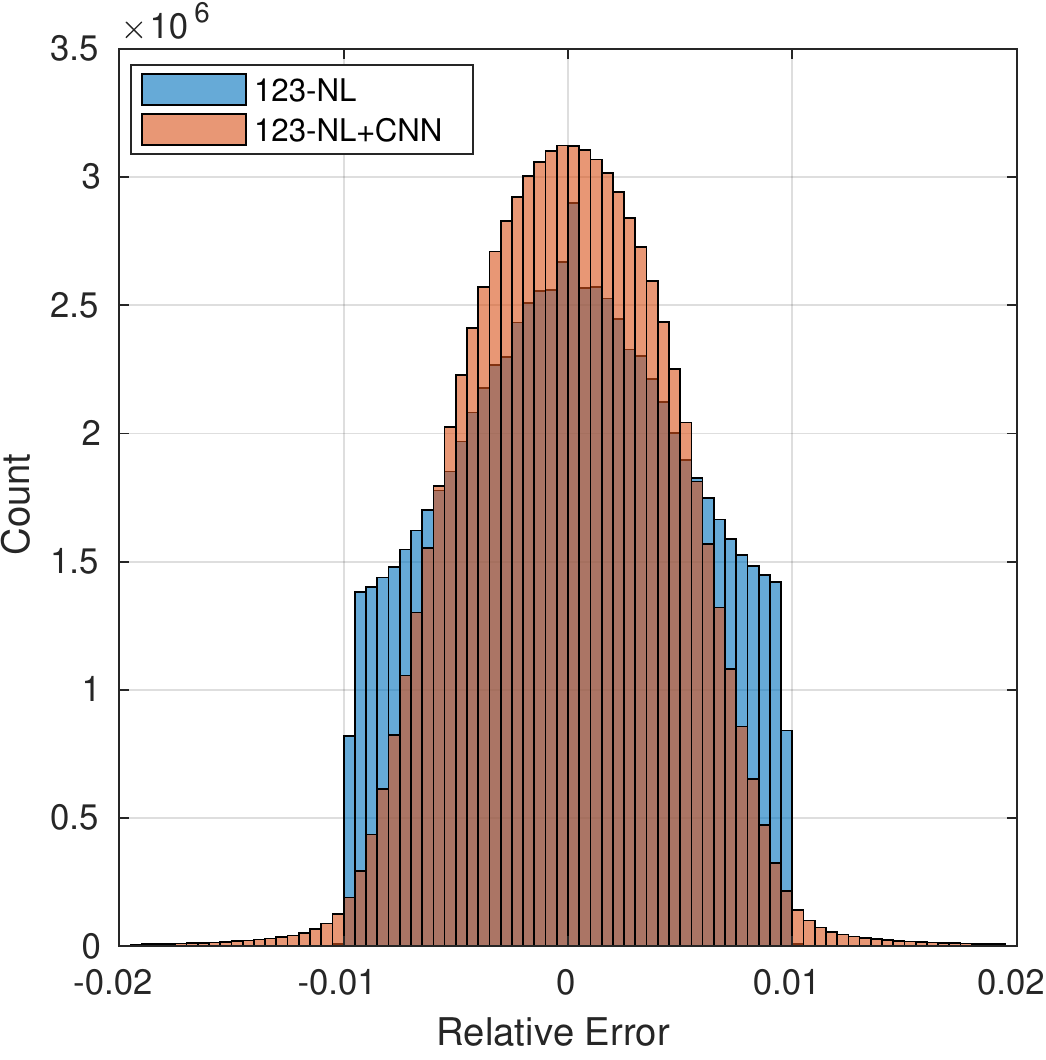}
    \caption{Lossy CCSDS 123.0-B-2}
    \end{subfigure}
    \begin{subfigure}{0.49\columnwidth}
    \includegraphics[width=\columnwidth]{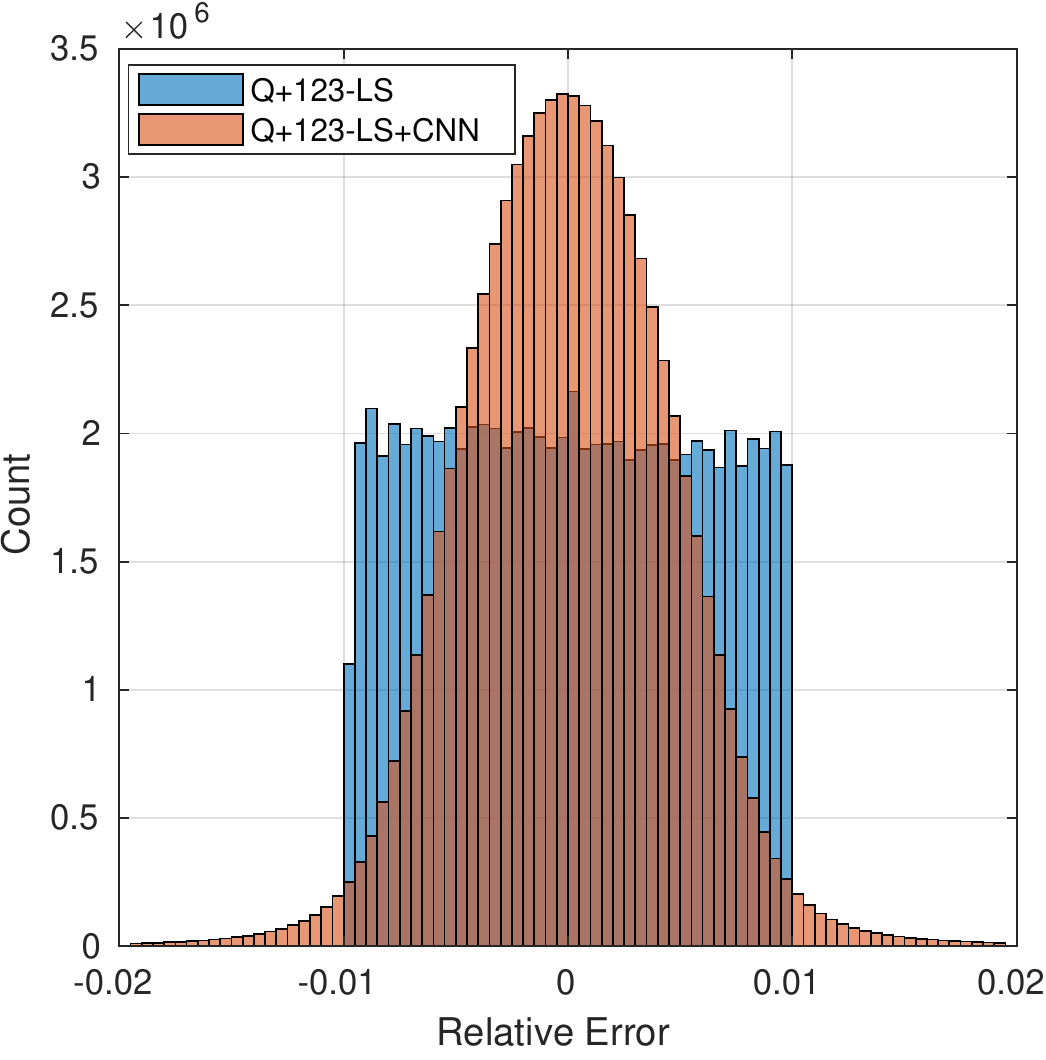}
    \caption{Prequantized}
    \end{subfigure}
    \caption{Error distribution for \textit{sc0} for $Q=61$.}
    \label{fig:error}
\end{figure}

\begin{figure}[t]
    \centering
    \begin{subfigure}{\columnwidth}
    \includegraphics[width=\columnwidth]{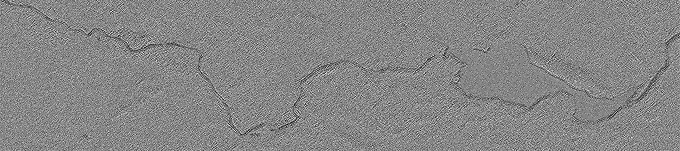}
    \caption{Lossy CCSDS 123.0-B-2 (CNN gain: 0.88 dB)}
    \end{subfigure}
    
    \begin{subfigure}{\columnwidth}
    \includegraphics[width=\columnwidth]{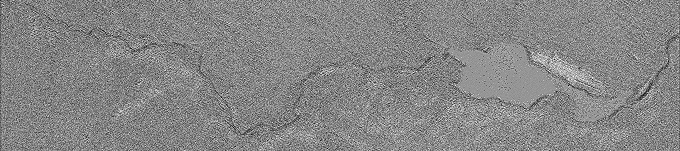}
    \caption{Prequantized (CNN gain: 1.13 dB)}
    \end{subfigure}
    \caption{CNN reconstruction residual $\mathbf{I}^{DQ} - \mathbf{I}^{Q}$ for $Q=31$. \textit{sc0} image, rows 150-300, all columns, band 47.}
    \vspace{-0.6cm}
    \label{fig:residual}
\end{figure}

The CNN described in Sec. \ref{sec:cnn} is trained from scratch with patches from scenes acquired by the target sensor. The number of patches should be large enough to represent the variability in the acquired scenes. Patches, instead of full scenes, can be used since the CNN is learning the distortion introduced by the compression process, which is local in nature. Once trained, the CNN can be used to restore any new scene acquired by that sensor without further fine-tuning. In a real operating scenario, one may not have realistic training data to begin with, e.g., just after the launch of the satellite. This can be easily solved by downloading a few scenes with lossless compression as one of the first tasks after deployment, and train the neural network using those (their compressed versions at different quality points can be easily produced by running the compression algorithm directly on the ground).

In our experiments, the CNN has been trained using 70000 patches of size $32 \times 32 \times 8$ randomly extracted from AVIRIS images from the Cuprite, Jasper and Moffett scenes. Notice that these are older scenes and have some artifacts with respect to newer scenes, showing that the proposed CNN is also robust to perturbations and that the overall performance could be further improved with a higher quality training set. Nevertheless, we used them as they are well-known and readily available to create a training set with sufficiently varied scenes. Patches have been extracted from the decoded images. Concerning the experiments on bounded absolute error, the following quantization step sizes have been chosen: $Q \in \lbrace 3,7,11,15,21,31,41,61,101 \rbrace$ for both the CCSDS and prequantization compressors to let the networks operate at roughly the same quality point. On the other hand, the following maximum absolute relative errors ( defined as $R=\max_{x,y,z} \frac{\vert I^{Q}_{x,y,z} - I_{x,y,z} \vert}{I_{x,y,z}}$), have been chosen for the experiments on bounded relative error: $R \in \lbrace 0.01,0.001,0.0075,0.005,0.0025,0.0005 \rbrace$. An independent model has been trained for each value of $Q$ and $R$ and each compression method. The clipping layer in the CNN implements the following operation
\begin{align*}
    E_{x,y,z}^{\text{CLIP}} = \begin{cases}
    -\Delta \quad &\text{if } E_{x,y,z} \leq -\Delta\\
    \Delta \quad &\text{if } E_{x,y,z} \geq \Delta\\
    E_{x,y,z} \quad &\text{otherwise}
    \end{cases},
\end{align*}
for the bounded absolute value experiments, and the following
\begin{align*}
    E_{x,y,z}^{\text{CLIP}} = \begin{cases}
    -RI^Q_{x,y,z} \quad &\text{if } E_{x,y,z} \leq -RI^Q_{x,y,z}\\
    RI^Q_{x,y,z} \quad &\text{if } E_{x,y,z} \geq RI^Q_{x,y,z}\\
    E_{x,y,z} \quad &\text{otherwise}
    \end{cases},
\end{align*}
for the bounded relative error experiments. As a remark, one might wonder why using an additive residual also for the reconstruction problem with bounded relative error, instead of a multiplicative residual: we found that a multiplicative residual caused instability in the training process.
We used the Adam optimization algorithm \cite{kingma2014adam} with a learning rate equal to $10^{-8}$ for a total number of iterations corresponding to 1000 epochs. It was noticed that models for small values of $Q$ and $R$ especially benefited from the low learning rate. The convolutional layers have a fixed number of filters equal to 64.
The CCSDS predictor has been set to use 3 prediction bands for both the lossy compressor and the lossless prediction after prequantization. 

The testing dataset is strictly disjoint from the training data and it is composed of the \textit{sc0}, \textit{sc3}, \textit{sc10}, \textit{sc11}, \textit{sc18} scenes from the AVIRIS Yellowstone images. We remark that these images have not been used during the training phase. For testing purposes the input to the network is a slice of the image with 8 bands and full spatial resolution ($512 \times 680 \times 8$). All the possible slices of 8 bands out of the available 224 bands are fed to the network by moving the window selecting the bands by one band at a time and finally merging the resulting images with a weighted average of the overlapped parts. Reconstructing one full image of size $512 \times 680 \times 224$ takes 64 seconds on an Nvidia GTX 1080 Ti with a peak GPU memory utilization of 4096 MB.
A C-language reference implementation of the CCSDS standard has been used to generate compression results while the CNN has been implemented with the PyTorch library. Code and pretrained models are available online\footnote{https://github.com/diegovalsesia/hyperspectral-dequantization}.

\subsection{Bounded absolute error}

\begin{figure*}[t]
    \centering
    \begin{subfigure}{0.48\textwidth}
    \includegraphics[width=\textwidth]{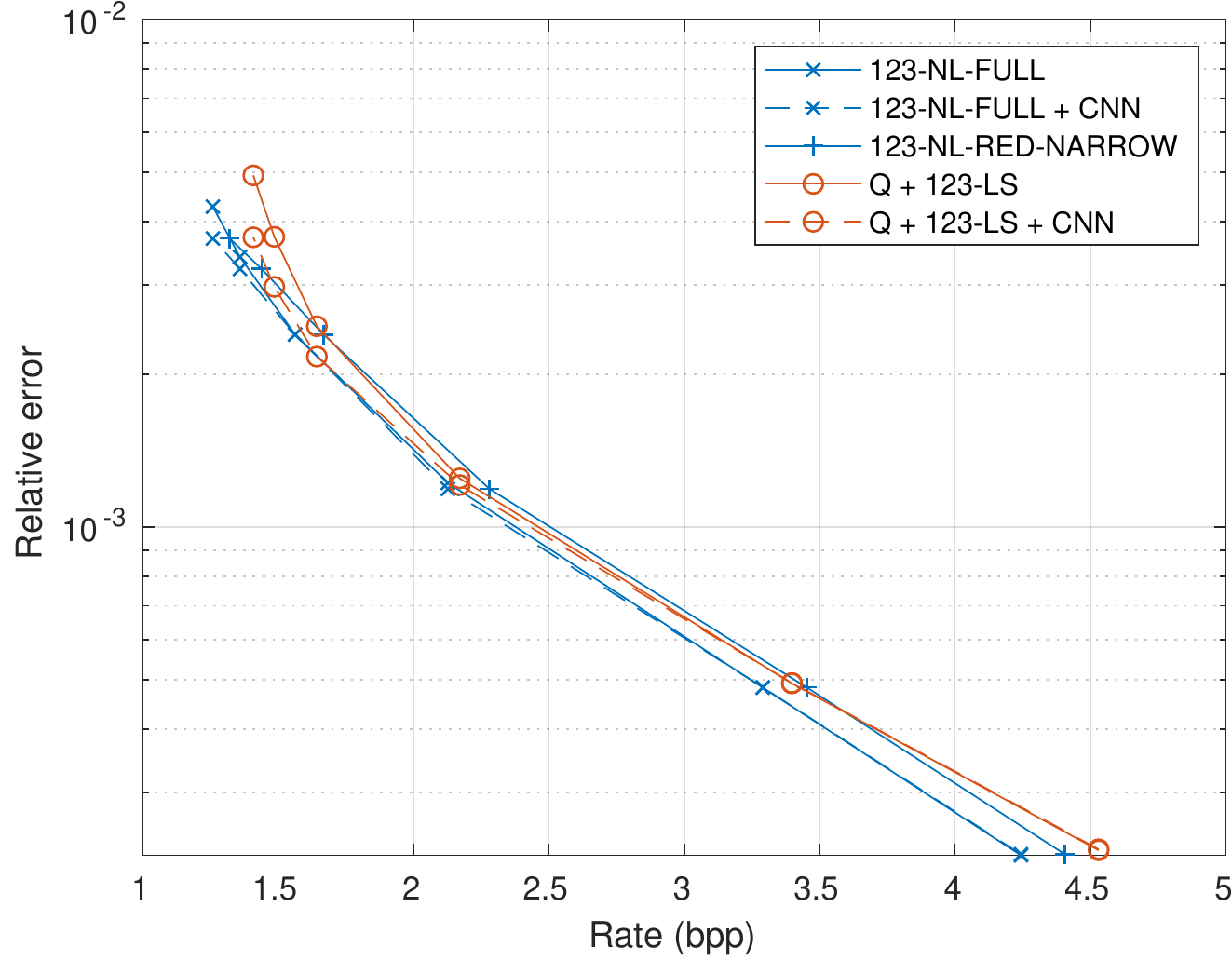}
    \caption{sc0}
    \end{subfigure}
    \begin{subfigure}{0.48\textwidth}
    \includegraphics[width=\textwidth]{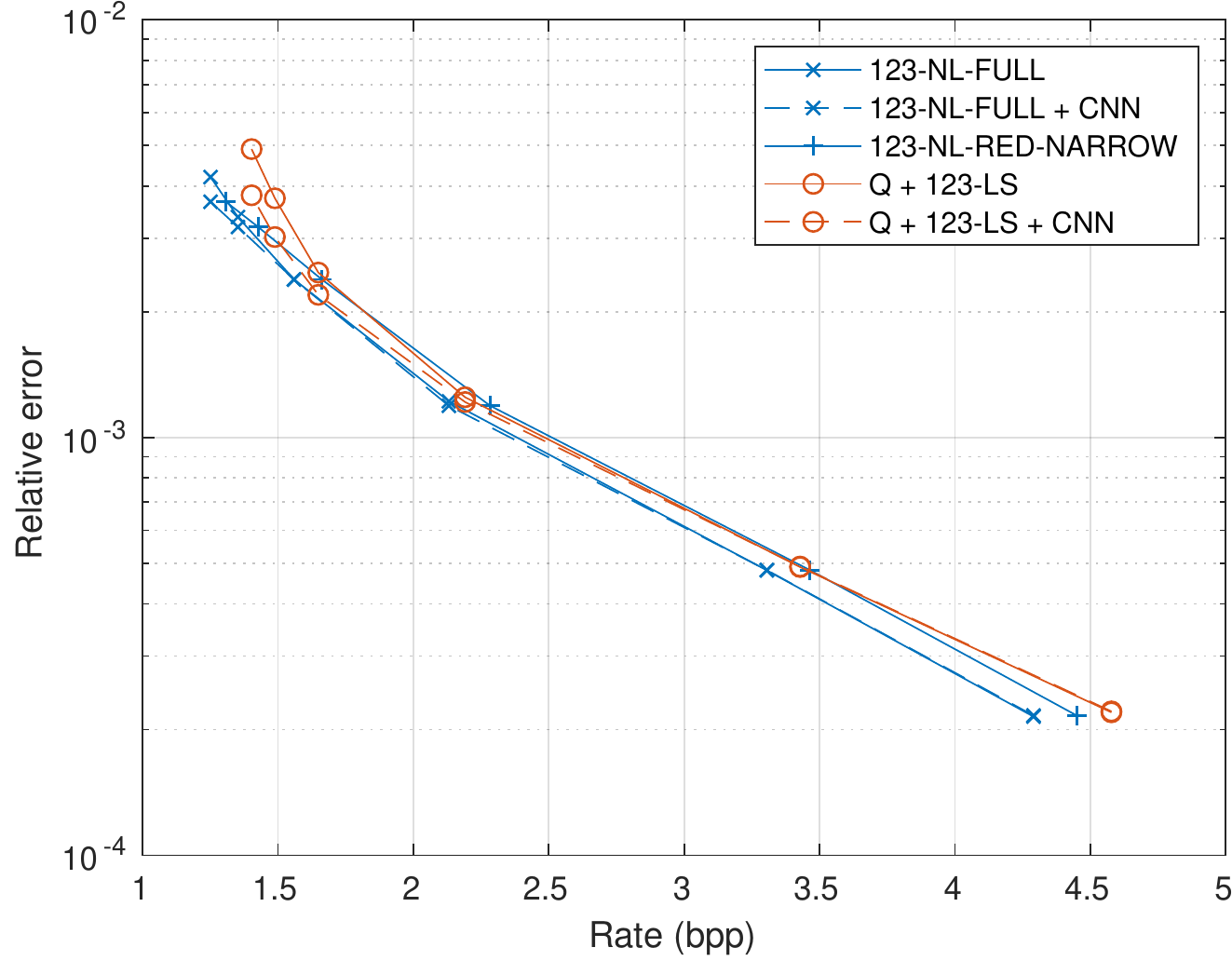}
    \caption{sc3}
    \end{subfigure}
    \begin{subfigure}{0.48\textwidth}
    \includegraphics[width=\textwidth]{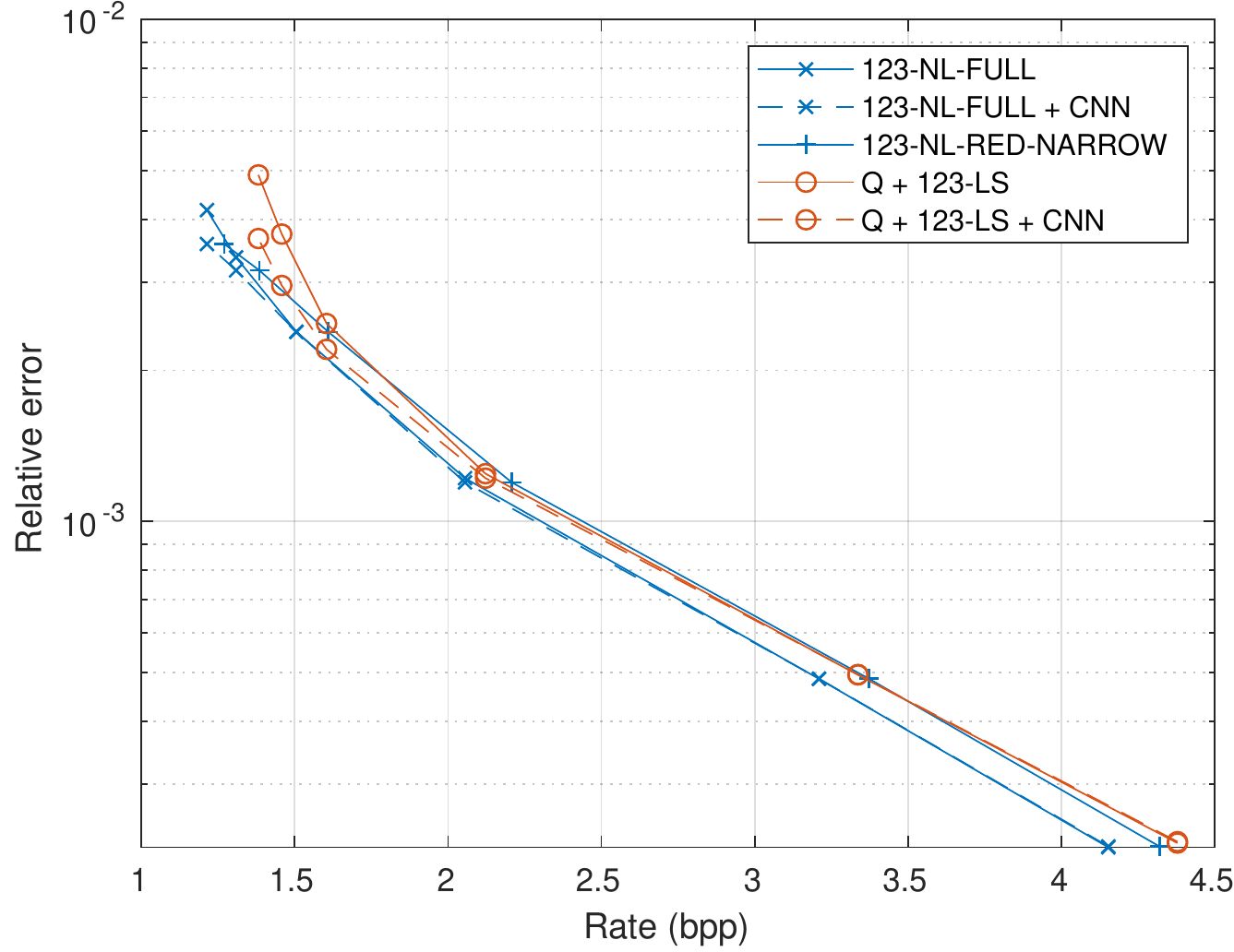}
    \caption{sc11}
    \end{subfigure}
    \begin{subfigure}{0.48\textwidth}
    \includegraphics[width=\textwidth]{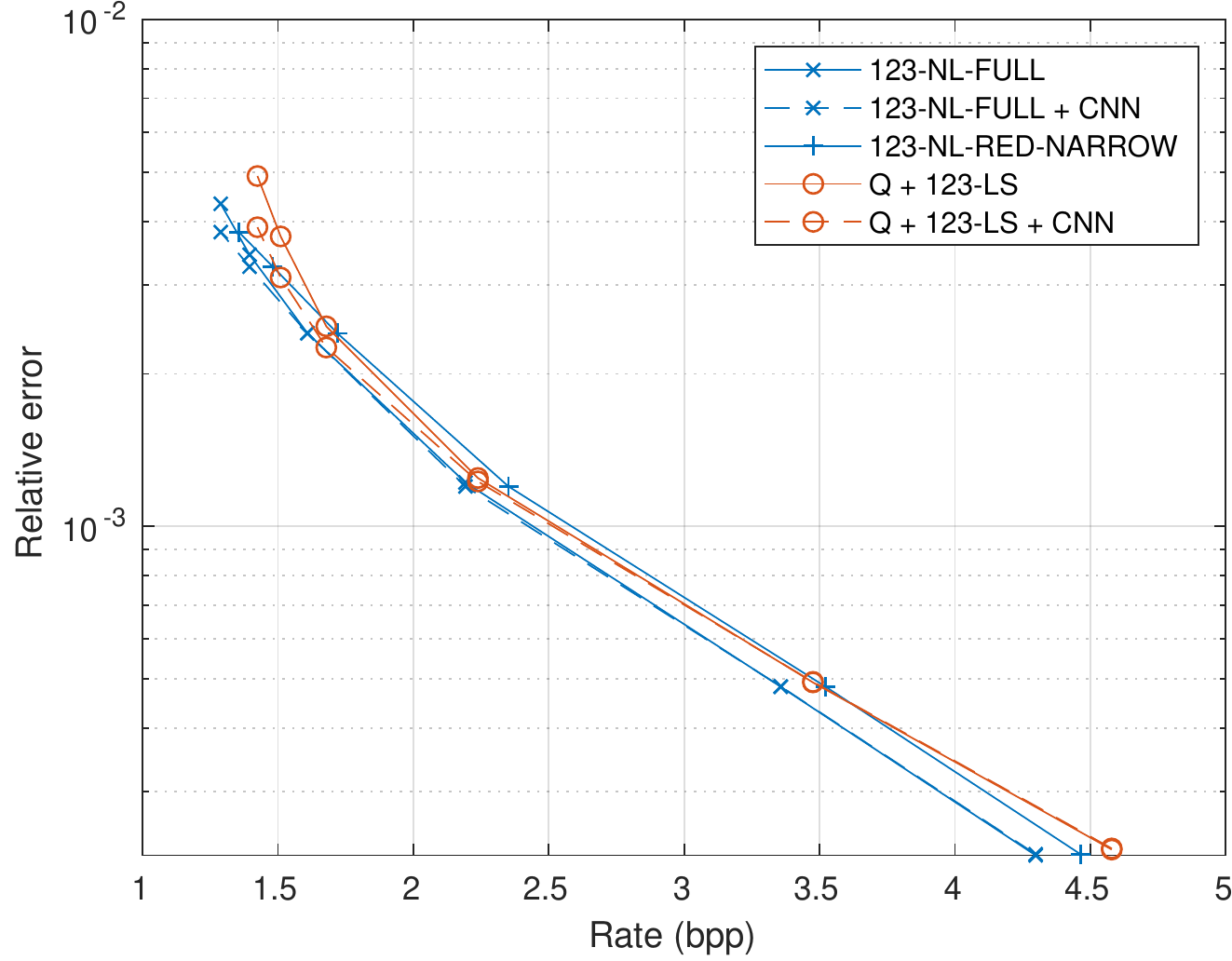}
    \caption{sc18}
    \end{subfigure}
    \caption{Rate-MARE performance of various compression methods with and without onground CNN. 123-NL: lossy CCSDS 123.0-B-2 (full, wide, neighbor-oriented mode); Q+123-LS: prequantization followed by lossless CCSDS 123.0-B-2 (full, wide, neighbor-oriented mode); 123-NL-RED-NARROW: lossy CCSDS 123.0-B-2 (reduced, narrow, neighbor-oriented mode); CNN: CNN reconstruction.}
    \vspace{-0.2cm}
    \label{fig:snr_bre}
\end{figure*}

\begin{table*}[]
\normalsize
\centering
\caption{Percentage mean absolute relative error for test set}
\label{table:mare}
\begin{tabular}{cccccc}
& \textbf{123-NL} & \textbf{123-NL + CNN} & \textbf{Q + 123-LS} & \textbf{Q + 123-LS + CNN} & \hspace{-9pt} \textbf{123-NL-RED-NARROW} \hspace{-9pt} \\ \hline
\textbf{1.5 bpp} & $(0.258 \pm 0.023)\%$            & $(\mathbf{0.255} \pm 0.020)\%$                  & $(0.348 \pm 0.033)\%$                & $(0.284 \pm 0.027)\%$                      & $(0.288 \pm 0.024)\%$                                                                                     \\ \hline
\textbf{2.0 bpp} & $(0.138 \pm 0.011)\%$            & $(\mathbf{0.136} \pm 0.012)\%$                  & $(0.153 \pm 0.012)\%$                & $(0.145 \pm 0.010)\%$                      & $(0.160 \pm 0.013)\%$                                                                                     \\ \hline
\textbf{3.0 bpp} & $(0.060 \pm 0.003)\%$            & $(\mathbf{0.060} \pm 0.003)\%$                  & $(0.066 \pm 0.004)\%$                & $(0.065 \pm 0.004)\%$                      & $(0.067 \pm 0.004)\%$                                                                                     \\ \hline
\textbf{4.0 bpp} & $(0.026 \pm 0.002)\%$            & $(\mathbf{0.026} \pm 0.002)\%$                  & $(0.032 \pm 0.003)\%$                & $(0.032 \pm 0.003)\%$                      & $(0.030 \pm 0.002)\%$                                                                                     \\ \hline
\end{tabular}
\end{table*}

The first experiment regards the rate-distortion performance of the two compressors and the relative gain provided by the CNN for the bounded absolute error scenario. Quality is measured by the SNR computed as
\begin{align*}
    \mathrm{SNR} = 10\log_{10} \frac{\sum_{i=1}^{N_\mathrm{pixel}} s_i^2}{\sum_{i=1}^{N_\mathrm{pixel}}(s_i-s^R_i)^2}.
\end{align*} 
Other metrics such as the maximum spectral angle and the average spectral angle have been studied in the literature \cite{christophe2005quality}, but we omit them as they follow the same trends observed for SNR.
Fig. \ref{fig:snr_bae} shows the rate-SNR curves for four test scenes. Table \ref{table:snr} reports the average SNR over the test set achieved by the various methods at four fixed rates (SNR values are linearly interpolated from the two closest available rate-distortion points). First, it can be noticed that the CNN provides more than 1 dB of improvement at 1.5 bpp, around 0.5 dB at 2.0 bpp and very small gains at high rates. Then, it is very interesting to notice that the sub-optimality of the prequantized method is quite limited and can be fully recovered by the CNN at all rates above or equal to 2.0 bpp. We also notice that the prequantized method is always better than lossy CCSDS 123.0-B-2 in reduced mode with narrow, neighbor-oriented local sums, which enables higher-throughput implementations, even without the help of the CNN. 

Fig. \ref{fig:error} shows the distribution of the error between the original \textit{sc0} image, the compressed version and the reconstructed version using the CNN for $Q=61$, for both compression techniques. It can be noticed that the CNN is able to reduce the average error amplitude, explaining the excess distribution around zero. We can also notice the longer tail of the error for the reconstructed image which is due to the ability to only guarantee twice the original bound after the reconstruction process, as explained in Sec. \ref{sec:cnn}. Fig. \ref{fig:residual} visually shows the residual correction, i.e., $\mathbf{E}^{\text{CLIP}} = \mathbf{I}^{DQ}-\mathbf{I}^Q$, estimated by the network to restore the image. We can notice that the action of the CNN is particularly significant around edges.

Finally, we remark that we also tested total variation regularization as defined in Eq. \eqref{eq:tv_min} but the gain was limited to 0.1 dB at 1.5 bpp, 0.05 dB at 2 bpp and no gain was observed at higher rates, for both compression techniques. This confirms that CNNs are able to exploit much more complex models to regularize the reconstruction problem.

\subsection{Bounded relative error}

\begin{figure}[t]
    \centering
    \begin{subfigure}{0.49\columnwidth}
    \includegraphics[width=\columnwidth]{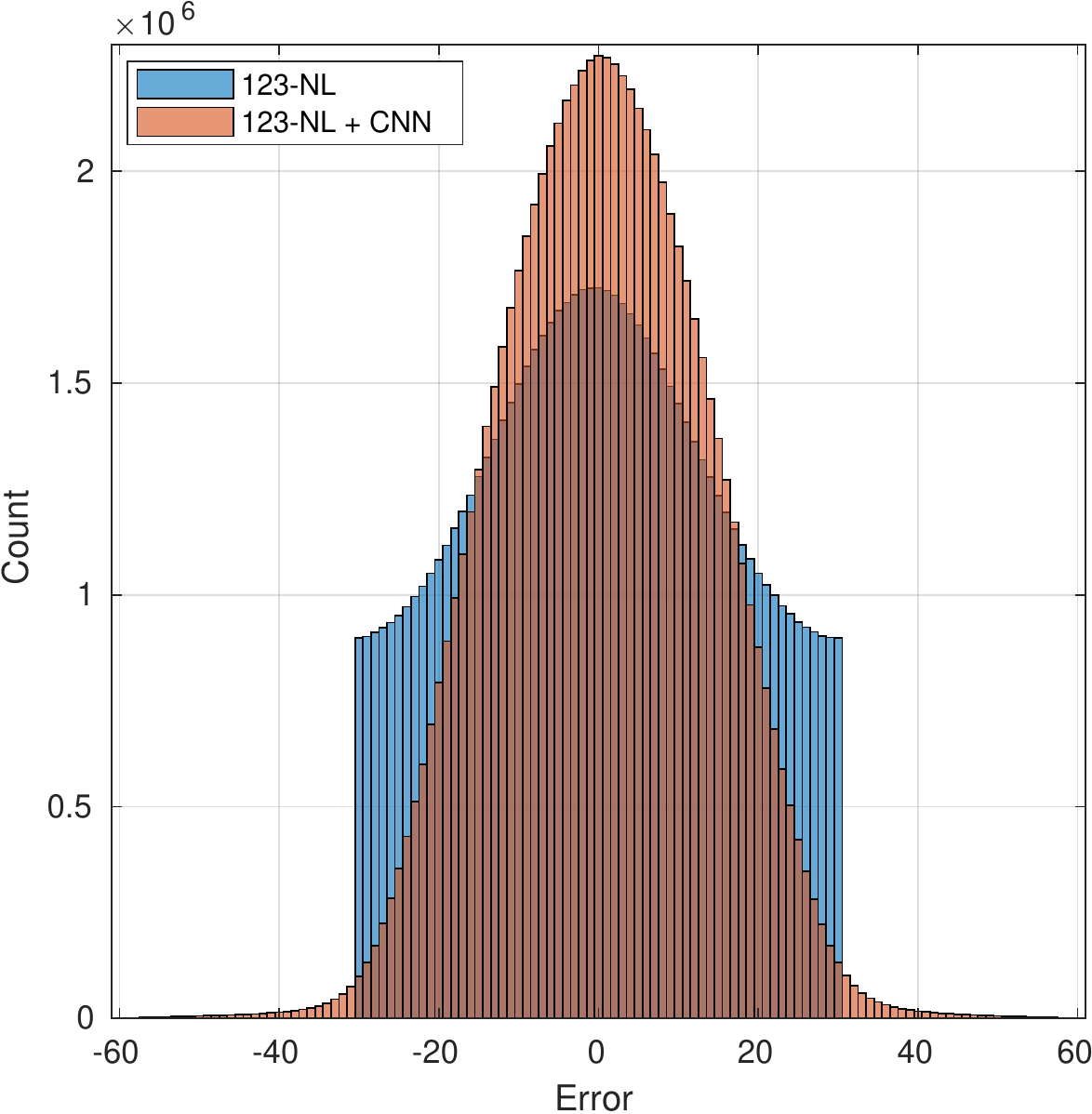}
    \caption{Lossy CCSDS 123.0-B-2}
    \end{subfigure}
    \begin{subfigure}{0.49\columnwidth}
    \includegraphics[width=\columnwidth]{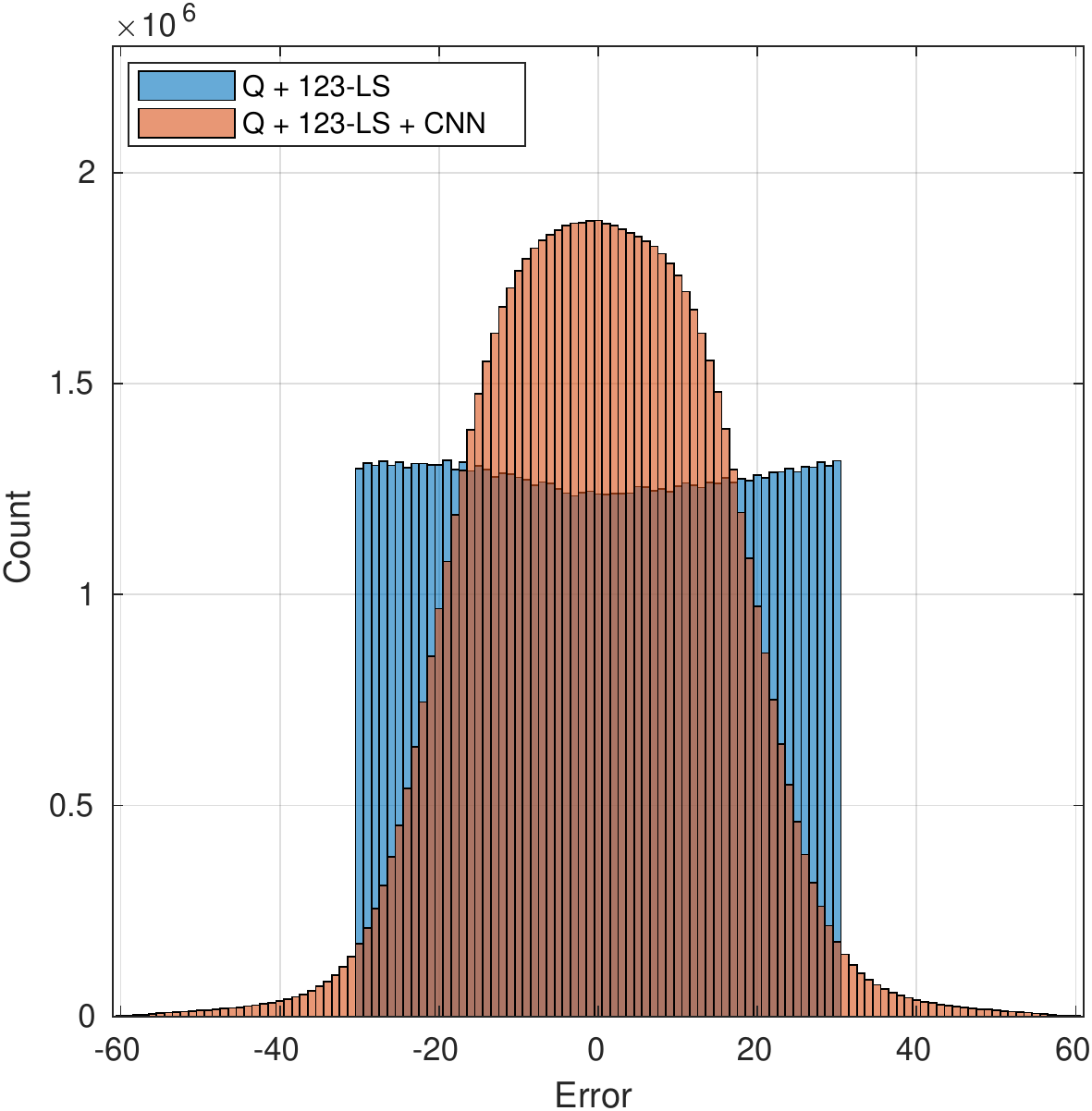}
    \caption{Prequantized}
    \end{subfigure}
    \caption{Relative error distribution for \textit{sc0} for $R=0.01$.}
    \label{fig:rel_error}
\end{figure}

In the experiments on bounded relative error we measure image quality in terms of mean absolute relative error (MARE) defined as:
\begin{align*}
    \mathrm{MARE} = \frac{1}{N_\mathrm{pixel}} \sum_{i=1}^{N_\mathrm{pixel}} \frac{\vert s_i-s^R_i \vert}{s_i}.
\end{align*} 
Fig. \ref{fig:snr_bre} shows the MARE as function of the rate for some test scenes. Table \ref{table:mare} also reports the achieved MARE for the different methods at fixed rate points. It can be noticed that CCSDS 123.0-B-2 in full, wide, neighbor-oriented mode followed by the CNN is confirmed as the best method. However, the gain provided by the CNN is quite limited with respect to the absolute error case. This may be due to the more challenging error statistics, being dependent on the signal in a multiplicative way. The prequantization method followed by the CNN is competitive with the CCSDS 123.0-B-2 full, wide, neighbor-oriented baseline, and can outperform the fast CCSDS 123.0-B-2 reduced, narrow, neighbor-oriented method. 
Fig. \ref{fig:rel_error} reports the relative error distribution with and without the CNN, again showing an excess around zero thanks to the CNN and a tail extending to twice the original maximum error target.

\subsection{Transfer learning experiment}
The optimal reconstruction results from the CNN can be obtained when the network is trained on images generated by the same sensor, so that the specific spatial and spectral correlation patterns or artifacts generated by that instrument can be exploited. However, the CNN works as a feature extractor and some of the features may generalize to different sensors. Table \ref{table:transfer} reports the results obtained by using the same CNNs trained from the AVIRIS images on the \textit{gran9} scene from the AIRS ultraspectral instrument, for the bounded absolute error mode. The size of this scene is equal to $135 \times 90 \times 1501$, thus having lower spatial resolution but higher spectral resolution with respect to the AVIRIS scenes. The results show that the CNNs perform well even if not trained specifically for the AIRS instrument. 

\begin{table*}[]
\normalsize
\centering
\caption{Transfer learning on AIRS sensor}
\label{table:transfer}
\begin{tabular}{cc|cc|cc}
\multicolumn{2}{c|}{\textbf{Q}}                & \textbf{123-NL} & \textbf{123-NL + CNN} & \textbf{Q + 123-LS} & \textbf{Q + 123-LS + CNN} \\ \hline
\multirow{2}{*}{3}   & SNR (dB)  & 68.73           & 68.83                 & 68.73               & 68.82                     \\   
                     & Rate (bpp) & \multicolumn{2}{c|}{2.81}                & \multicolumn{2}{c}{2.87}                        \\ \hline
\multirow{2}{*}{7}   & SNR (dB)  & 60.96           & 61.25                 & 60.95               & 61.45                     \\   
                     & Rate (bpp) & \multicolumn{2}{c|}{1.87}                & \multicolumn{2}{c}{1.99}                        \\ \hline
\multirow{2}{*}{11}  & SNR (dB)  & 57.07           & 58.15                 & 56.97               & 58.03                     \\   
                     & Rate (bpp) & \multicolumn{2}{c|}{1.54}                & \multicolumn{2}{c}{1.68}                        \\ \hline
\multirow{2}{*}{15}  & SNR (dB)  & 54.53           & 56.26                 & 54.26               & 55.53                     \\   
                     & Rate (bpp) & \multicolumn{2}{c|}{1.39}                & \multicolumn{2}{c}{1.54}                        \\ \hline
\multirow{2}{*}{21}  & SNR (dB)  & 51.97           & 53.91                 & 51.32               & 53.49                     \\   
                     & Rate (bpp) & \multicolumn{2}{c|}{1.25}                & \multicolumn{2}{c}{1.43}                        \\ \hline
\multirow{2}{*}{31}  & SNR (dB)  & 49.21           & 51.59                 & 47.94               & 52.01                     \\   
                     & Rate (bpp) & \multicolumn{2}{c|}{1.14}                & \multicolumn{2}{c}{1.33}                        \\ \hline
\multirow{2}{*}{41}  & SNR (dB)  & 47.16           & 49.51                 & 45.51               & 48.88                     \\   
                     & Rate (bpp) & \multicolumn{2}{c|}{1.10}                & \multicolumn{2}{c}{1.28}                        \\ \hline
\multirow{2}{*}{61}  & SNR (dB)  & 44.11           & 46.94                 & 42.05               & 46.85                     \\   
                     & Rate (bpp) & \multicolumn{2}{c|}{1.06}                & \multicolumn{2}{c}{1.23}                        \\ \hline
\multirow{2}{*}{101} & SNR (dB)  & 40.09           & 41.42                 & 37.66               & 43.00                     \\   
                     & Rate (bpp) & \multicolumn{2}{c|}{1.04}                & \multicolumn{2}{c}{1.17}                        \\ \hline
\end{tabular}
\end{table*}

\section{Conclusions} \label{sec:conclusions}
We proposed a method to compress hyperspectral images composed of an onboard predictive compressor and a ground-based CNN to reconstruct the decoded images and analyzed how it relates to the new CCSDS-123.0-B-2 recommendation. We showed that an onboard component based on prequantization followed by the lossless mode of CCSDS-123.0-B-2 can be significantly faster than the lossy mode of the standard and that, when coupled with the onground CNN, the same rate-distortion performance of the most efficient mode of lossy CCSDS-123.0-B-2 is achieved.

\end{document}